\documentclass[10pt,journal,compsoc]{IEEEtran}
\usepackage{amsmath}
\usepackage{bm}
\usepackage{listings}
\usepackage{booktabs}
\usepackage{amsfonts}
\usepackage{color}
\usepackage{graphicx}
\usepackage{enumitem}
\usepackage{multirow}
\usepackage{amsmath}
\usepackage{float}
\usepackage{cite}
\usepackage{subcaption}
\usepackage[ruled,vlined,linesnumbered]{algorithm2e}

\usepackage{enumerate}
\usepackage{amssymb}
\usepackage{amsthm}

\theoremstyle{remark}
\usepackage{mathtools}

\usepackage[table]{xcolor}
\def\hlb{}

\begin{document}
\title{When Attackers Meet AI: Learning-empowered Attacks in Cooperative Spectrum Sensing}

\author{Zhengping~Luo, Shangqing~Zhao,
        Zhuo~Lu, Jie~Xu, ~and~Yalin~E.~Sagduyu
\IEEEcompsocitemizethanks{
\IEEEcompsocthanksitem Zhengping Luo, Shangqing Zhao and Zhuo Lu are with the Department of Electrical Engineering, University of South Florida, Tampa FL, 33620. Emails: \{zhengpingluo, shangqing, zhuolu\}@usf.edu.
\IEEEcompsocthanksitem Jie Xu is with Department of Electrical and Computer Engineering
University of Miami, Coral Gables, FL 33146. Email: jiexu@miami.edu.
\IEEEcompsocthanksitem Yalin E. Sagduyu is with Intelligent Automation Inc, 15400 Calhoun Dr \#190, Rockville, MD 20855. Email: ysagduyu@i-a-i.com.}
}

\IEEEtitleabstractindextext{%
\begin{abstract}
Defense strategies have been well studied to combat Byzantine attacks that aim to disrupt cooperative spectrum sensing by sending falsified versions of spectrum sensing data to a fusion center. However, existing studies usually assume network or attackers as passive entities, e.g., assuming the prior knowledge of attacks is known or fixed. In practice, attackers can actively adopt arbitrary behaviors and avoid pre-assumed patterns or assumptions used by defense strategies. In this paper, we revisit this security vulnerability as an adversarial machine learning problem and propose a novel learning-empowered attack framework named Learning-Evaluation-Beating (LEB) to mislead the fusion center. Based on the black-box nature of the fusion center in cooperative spectrum sensing, our new perspective is to make the adversarial use of machine learning to construct a surrogate model of the fusion center's decision model. We propose a generic algorithm to create malicious sensing data using this surrogate model. Our real-world experiments show that the LEB attack is effective to beat a wide range of existing defense strategies with an up to 82\% of success ratio. Given the gap between the proposed LEB attack and existing defenses, we introduce a non-invasive method named as influence-limiting defense, which can coexist with existing defenses to defend against LEB attack or other similar attacks. We show that this defense is highly effective and reduces the overall disruption ratio of LEB attack by up to 80\%.
\end{abstract}

}

\maketitle

\IEEEdisplaynontitleabstractindextext

\IEEEpeerreviewmaketitle

\IEEEraisesectionheading{\section{Introduction}\label{sec:introduction}}

\IEEEPARstart{C}{ooperative} spectrum sensing has been proposed as an effective mechanism to enhance the spectrum sensing performance using cognitive radio devices (e.g., TV-band devices coded in IEEE 802.22). It enables a data fusion-based decision framework, in which multiple nodes report their sensing results to a fusion center that makes a centralized decision to enhance the spectrum sensing accuracy. This, however, opens up opportunities to Byzantine attacks (also widely referred to as spectrum sensing data falsification (SSDF) attacks) \cite{zhang2015byzantine, li2010catch, chen2008robust, kaligineedi2010malicious, rawat2011collaborative, fatemieh2011using, wang2014secure, yan2012vulnerability}, in which attackers aim to send malicious sensing results to fool the fusion center into making wrong decisions on the channel availability.

Given Byzantine attacks, defense mechanisms in cooperative spectrum sensing have been widely studied and include: (i) statistics-based mechanisms, which aim to build statistical models to detect or eliminate attackers \cite{kaligineedi2010malicious, chen2017cooperative, chen2012robust, penna2012detecting}; (ii) machine learning-based mechanisms, in which machine learning methods are used as countermeasures \cite{thilina2013machine, fatemieh2011using, wang2018primary, li2010catch, rajasegarar2015pattern}; (iii) trust (or reputation)-based mechanisms, i.e., building trust metrics for each node such that it can be weighted in the decision process \cite{chen2008robust,rawat2011collaborative, sagduyu2014trust}.

Many of existing defenses assume network or attackers as passive entities, e.g., assuming attackers will employ a pre-assumed or fixed attack strategy. Most of them have been demonstrated to be practical and effective under specific scenarios. For example, methods used in \cite{kaligineedi2010malicious,chen2017cooperative,penna2012detecting} assume that attacks behave in a particular way or prior information of attack statistics is known  such that a statistical model of an attack can be effectively built; and machine learning-based methods \cite{fatemieh2011using, wang2018primary, li2010catch, rajasegarar2015pattern} assume that malicious data pattern deviates from normal data pattern under a given classification rule. These methods are effective and often the best choice for the fusion center given some pre-assumed attack strategies. However, more advanced defenses are required to counter intelligent learning-based attacks (i.e., active attackers), where attack strategies are unknown \emph{a priori}.

Attackers can try to actively avoid pre-assumed behaviors or break assumptions used by defense mechanisms. Moreover, the time-varying nature of wireless channels and signals can deviate data properties of the legitimate sensing results from time to time. In this regard, pre-trained statistical models used by defense mechanisms may face a model mismatch phenomenon over time, which can be further exploited by attackers. All these observations motivate us to rethink Byzantine attacks from the new perspective of adversarial machine learning.


From the attackers' point of view, the fusion center of cooperative spectrum sensing has three main characteristics: (i) sensing nodes report their sensing data as inputs to the fusion center; (ii) the fusion center announces the final channel status decision as the output; and (iii) the fusion center tries to make better decisions compared to independent sensing decisions of each individual sensing node.

\begin{figure}[t]
\centering
\includegraphics[scale=0.8]{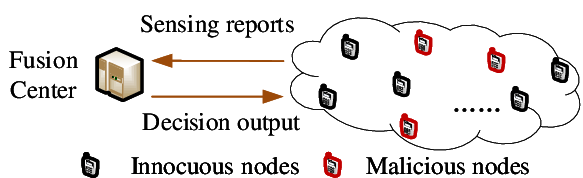}
\caption{System model.}
\label{Fig:system_model}
\end{figure}

We treat the fusion center, which consists of defenses and cooperative sensing decision rules, as a black box with known inputs and outputs, as illustrated in Fig.~\ref{Fig:system_model}. The third characteristic pointed out above opens the door to use a small number of nodes to approximate the decision model at the fusion center.

Inspired by \textit{no free lunch} theorems \cite{wolpert1997no} and the \textit{transferability} \cite{papernot2016transferability} property in machine learning, attackers can, in fact, use inputs and outputs shown in Fig.~\ref{Fig:system_model} to build an approximate model, also called as surrogate model, of the targeted fusion center (as an exploratory (or inference) attack), and then launch effective attacks with minimum data manipulation to mislead the fusion center.

The surrogate model is essentially a partial model of the fusion center. To make the partial model attack on the cooperative spectrum sensing process, we propose a \textit{Learning-Evaluation-Beating (LEB)} attack framework in this paper. The LEB attack consists of three steps:  (i) the attacker first learns  its own surrogate model to approximate the fusion center; (ii) the attacker evaluates whether the learned model is accurate enough or not to launch attacks; and (iii) the attacker combines the sensed data with a malicious perturbation as the input to the fusion center.

We also propose a learning algorithm based on a set of sub-models and a generic data generation algorithm to generate adversarial examples (i.e., falsified data with the minimum data manipulation to flip the fusion center's decision) in sub-models. We present comprehensive real-world experiments to measure the performance of spectrum sensing under LEB attacks and under a wide range of existing defense methods.

When an LEB attacker is present in cooperative spectrum sensing, it is challenging to build an effective defense mechanism to distinguish LEB attacker or similar malicious nodes from normal nodes. The principal reason for the success of LEB attacker is that they can build up their influence or impact on the fusion center by taking advantage of the model mismatch phenomenon. We introduce a new metric named  decision-flipping influence to measure the influence of a given subset of sensing nodes.

Furthermore, we propose an influence-limiting defense to evaluate and limit the influence that any subset of nodes have on the fusion center, thus decreasing or eliminating attack capabilities. We design the influence-limiting defense as a make-up to existing defenses such that the performance of fusion center against the learning-empowered attack can become more robust. Our experimental results demonstrate that the proposed defense can effectively bridge the gap between traditional defenses in cooperative spectrum sensing and new LEB attacks.

Our major contributions are listed as follows:
\begin{itemize}
\item We rethink traditional security vulnerabilities of cooperative spectrum sensing and present a new adversarial machine learning perspective to create a powerful attack mechanism named LEB attack, and show that the traditional duel of attacks and defenses in cooperative spectrum sensing has to be re-visited in the presence of new learning-based attack models.
\item We introduce a make-up mechanism named influence-limiting defense to existing defenses to counter LEB attacks or other similar learning-based attacks.
\item We collected real-world spectrum data to validate the model mismatch phenomenon. Our experiments show that the LEB attack can achieve an up to 82\% attack success ratio while only manipulating a small number of malicious nodes. Experimental results of influence-limiting defense demonstrate an average overall disruption ratio reduction by 78\% under LEB attacks compared with traditional defenses.
\end{itemize}
Other contributions include the following: the surrogate model is designed in a flexible way to adopt a wide range of sub-models; a generic adversarial sample generation algorithm based on the sub-model set is proposed to create samples against the fusion center; the LEB attack and influence-limiting defense proposed in this paper can be extended to a more generic partial model problem in machine learning, in which an attacker controls part of the input dimensions to compromise the machine learning model.

The rest of the paper is organized as follows. Section~\ref{sec:preliminary} introduces the  system  model,  background and model mismatch phenomenon in realistic network scenarios. Section~\ref{sec:LEB} describes the LEB attack framework. Experimental evaluation of the LEB attack framework is presented in Section~\ref{exper_eval}. Section~\ref{sec:defense} proposes a defense mechanism against LEB attack and other similar attacks. Limitations of the LEB attack framework, influence-limiting defenses and the future work are discussed in Section~\ref{sec:limitations}. Section~\ref{sec:related} overviews the related work and Section~\ref{sec:conclusion} concludes the paper.

\section{Model and Preliminaries}\label{sec:preliminary}
In this section, we introduce the system model, overview existing studies, and identify challenges.

\subsection{System Model}\label{system_model}
We consider a cooperative spectrum sensing scenario with $n$ sensing nodes and one fusion center, as shown in Fig.~\ref{Fig:system_model}. At each timeslot (i.e., each round of sensing), all nodes perform spectrum sensing on a TV spectrum channel, then report their results to the fusion center that makes the global channel usability decision based on all inputs. We assume that energy detection \cite{chen2017cooperative,kaligineedi2010malicious} is employed at each node. A sensing report contains the value of the energy level sensed by the corresponding node.

\begin{figure*}
\centering
\includegraphics[scale=0.8]{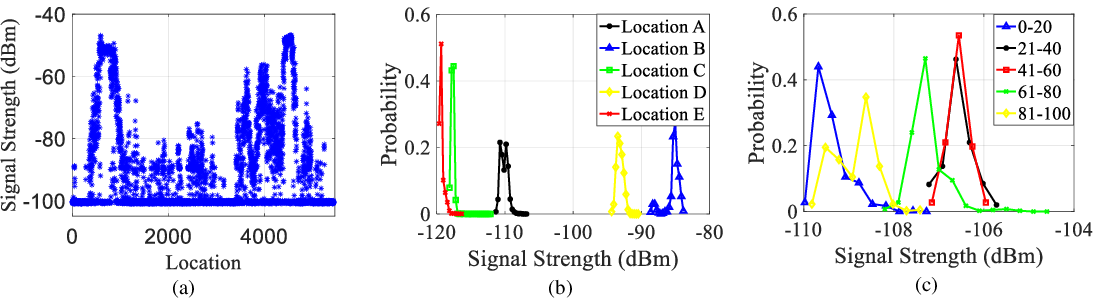}
\caption{(a) Signal strengths for one channel over 5282 different locations in Atlanta metropolitan area \cite{saeed2017local}; (b) Signal strengths distribution over different locations in our dataset; (c) Signal strengths distribution over different time periods (0-20, 21-40, 41-60, 61-80, 81-100) in our dataset.}
\label{Fig:model_mismatch}
\end{figure*}


The energy level vector at the $i$th timeslot from $n$ sensing nodes is denoted by $\textbf{x}_i = [x_{i,1},x_{i,2},...,x_{i,n}]^\top, \textbf{x}_i \in \mathcal{X}\subset \mathbb{R}^{n \times 1}$, where $[\cdot]^\top$ is the matrix transpose operator. $\mathcal{X}\subset \mathbb{R}^{n \times 1}$ is the value space for the sensed energy level, and $\mathbb{R}^{n \times 1}$ is an $n$-dimensional Euclidean space. Historical sensing results of all sensing nodes are denoted as $\{\textbf{h}_k\}_{k \in [1,n]} =[x_{1,k},x_{2,k},x_{3,k},...,x_{i,k}]^\top$.

We denote the decision mapping function (implemented by the data fusion rule and potential defenses) used at the fusion center as $O:\mathcal{X} \rightarrow \mathcal{Y}$, where $\mathcal Y = \{-1,1\}$ is the decision output space with $-1/1$ denoting the channel being available/unavailable. The fusion center makes the final channel status decision $y_i \in \mathcal{Y}$ based on sensed results $\textbf{x}_i$. We assume the attacker has no information about the decision mapping function adopted by the fusion center while it knows only the final decision $y_i$ at each timeslot $i$. 

Cooperative spectrum sensing enables multiple nodes to report their sensing results to a fusion center to enhance the spectrum sensing accuracy. In the meantime, Byzantine nodes can send malicious sensing results to the fusion center and mislead the fusion center to make wrong decisions about the channel status and further cause interference between primary users of the spectrum with secondary users.

Without loss of generality, we assume that the first $m$ (out of $n$) nodes are malicious nodes that are fully controlled by the attacker. The attacker can make malicious nodes to report whatever sensing results needed but has no information about the reports of remaining $n-m$ innocuous nodes.

We denote the sensed data vector by malicious nodes at timeslot $i$ as $ \mathbf{a}_i = [x_{i,1},x_{i,2},...,x_{i,m}]^\top, \mathbf{a}_i \in \mathcal{A} \subset \mathbb{R}^{m \times 1}$, where $\mathcal{A}$ is the report space of malicious nodes. Obviously, $\mathbf{a}_i$ is the first part of $\mathbf{x}_i$ that
\begin{equation}
\mathbf{x}_i = [\underbrace{x_{i,1},x_{i,2},...,x_{i,m}}_{\text{\textbf{a}}_i},x_{i,m+1},...,x_{i,n}]^\top, \textbf{x}_i \in \mathcal{X}.
\end{equation}

The objective of the attacker is to manipulate sensing data vector $\mathbf{a}_i$ to mislead the fusion center to yield a wrong sensing decision $y_i$.

\subsection{Defending against Byzantine Attacks} \label{defenses}
Defenses against Byzantine attacks in cooperative spectrum sensing can be essentially viewed as an anomaly detection problem, in which anomalous (malicious) nodes need to be distinguished from innocuous nodes. Many of defenses proposed in previous works correspond to a combination of two or  multiple aspects of node characteristics, such as transmit pattern and statistical consistency. Here, we classify these defenses into three broad categories by the main node characteristic used by defenses:
\begin{itemize}[leftmargin=*]
\item \textit{Statistics-based defense,} which assumes that the attack has a particular behavior or prior information of network or attack statistics are known \cite{kaligineedi2010malicious, chen2017cooperative,penna2012detecting}. As the most widely used approach to detect malicious nodes, statistics-based schemes defend fusion centers against attacks based on the statistical measures derived from the sensing report history $\{\textbf{h}_k\}_{k \in [1,n]}$, such as covariance and  deviation \cite{kaligineedi2010malicious,chen2017cooperative,chen2012robust}.

\item \textit{Machine learning-based defense,} in which malicious nodes are assumed to have different underlying data patterns, for example, historical sensing results $\{\textbf{h}_k\}_{k \in [1,m]}$ of malicious nodes and $\{\textbf{h}_k\}_{k \in [m+1,n]}$ of innocous nodes can be classified into different categories by machine learning techniques. Both supervised \cite{fatemieh2011using} and unsupervised \cite{wang2018primary, li2010catch, rajasegarar2015pattern} methods have been proposed to identify or eliminate the effects malicious nodes.

\item \textit{Trust (or reputation)-based defense,} the essence of which is to compute a trust metric based on $\{\textbf{h}_k\}_{k \in [1,n]}$. Those nodes with low trust metrics are detected as malicious nodes and get eliminated or less weighted from the decision process. It is worth noting that trust-based methods usually compute trust metrics based on mathematical deviations obtained from statistics-based mechanisms \cite{rawat2011collaborative,chen2008robust}.
\end{itemize}

These defense methods have been demonstrated to be effective and practical in countering attacks. However, their common assumption is that attackers are passive, i.e., they follow a fixed or pre-assumed attack strategy. New learning-empowered active attackers will have the opportunity to take advantage of these defenses. 

\subsection{Model Mismatch in Realistic Network Scenarios} \label{unevenness}

Given assumptions made by existing defenses against Byzantine attacks, the question is whether there is any model mismatch phenomenon between the assumed and actual (real-world) scenarios, i.e., whether the statistical properties and other measures like pattern and reputation for each node will differ, or not. We explore the model mismatch phenomenon both in spatial and temporal dimensions.

Two datasets are used in our exploration. The first one is a public dataset collected from 5282 different locations in Atlanta metropolitan area over 13 TV white space channels \cite{saeed2017local}. We use the dataset to explore the model mismatch phenomenon in spatial dimension. The other one is collected from a time period of 100 hours over 22 channels, 5 different locations were sensed over a 20 $\times$ 20 km$^2$ urban region. We use USRP N210 \cite{ettus2005usrp} as the sensor. This dataset is used to evaluate the model mismatch phenomenon in temporal dimension.

The model mismatch phenomenon in spatial dimension is illustrated in Fig.~\ref{Fig:model_mismatch}(a) and Fig.~\ref{Fig:model_mismatch}(b). Fig.~\ref{Fig:model_mismatch}(a) shows the signal strengths for one TV channel (randomly chosen from 13 channels) over 5282 different locations. The signal strength varies from -45dBm to -100dBm. Fig.~\ref{Fig:model_mismatch}(b) shows the probability distribution in terms of signal strengths of 5 locations from our collected dataset. It is obvious that locations C and E have the lowest signal strength while location B has the highest signal strength for the given channel.

The model mismatch phenomenon in temporal dimension is shown in Fig.~\ref{Fig:model_mismatch}(c). The 100-hour dataset is divided into 5 sub-datasets, each of which includes 20 hours of the data. The probability distributions of the signal strengths for each sub-dataset (averaged over 22 channels) are plotted in Fig.~\ref{Fig:model_mismatch}(c), which shows that the probability distribution changes over different sub-datasets. This observation indicates that a statistical or data model built for one time period changes for another time period.

Our data analysis on existing datasets and collected datasets demonstrate that the statistical/data model mismatch phenomenon over space and time is a real-world problem. In another words, it can be difficult for signal strengths of both malicious and innocuous nodes to follow exact models/behaviors assumed in a defense strategy. The existence of such phenomenon is also reasonable because of (i) environmental factors, such as weather and buildings, and (ii) network factors, such as co-channel interference and adjacent-channel leakages from other broadcasting activities.

The model mismatch phenomenon further leads to two consequences: (i) uncertainty of sensing results occurs among multiple nodes (even among innocuous nodes) simultaneously, and (ii) it will be challenging to find values of parameters for the optimal decision to decide the channel status over time. Thus, an intelligent attacker can try to learn decision models, then takes advantage of the learned model and generates malicious sensing data to mislead the fusion center.

\section{LEB Attack Framework: Motivation and Design} \label{sec:LEB}
In this section, we first present the motivation behind the adversarial learning-based attack design and then propose the LEB attack framework against cooperative spectrum sensing.

\subsection{Attack Motivation} \label{sec:attackmotivation}
As we can observe from the model mismatch phenomenon, it is often impractical for the fusion center to have perfect models about real-world signal data statistics or patterns due to the spatially and temporally varying nature of the wireless environment. Moreover, if we turn the table around and think from an intelligent attacker's perspective, the attacker should avoid any known behavior as assumed in traditional defenses and make its attack stealthy.

Our key observations for designing a new attack model are threefold: (i) network nodes always report their sensing data to the fusion center; (ii) \hlb{as spectrum sensing applications are built upon the wireless scenario and the goal of sensing is to let nodes know the availability of a spectrum band (e.g., for them to access the channel opportunistically), it is reasonable to assume that the final decision is broadcast by the fusion center to all nodes to inform them of the decision to use (or not use) the band} (iii) many of cooperative spectrum sensing methods \cite{chen2017cooperative,kaligineedi2010malicious,chen2008robust,li2010catch} are based on the fact that each node performs the channel sensing task independently and the fusion center tries to make a better decision given independent decisions from all nodes.

Based on the above three observations, we treat the fusion center as a black box such that the decision model (including both fusion rules and potential defenses) can be considered as a black-box function with known inputs and outputs. As a result, the attacker uses inputs and outputs of this back box to build a surrogate model of the decision model used at the targeted fusion center. The third observation opens the door to use a small number of nodes ($m < n$) to build a surrogate model. If $m = n$, then the attacker has full control of the fusion center and the attack success ratio will be 100\%.  When $m < n$, it will be difficult for the attacker to achieve 100\% of the attack success ratio, but there is still a probability to succeed. The design of the LEB attacker is based on this point. After stealing the decision model, the attacker can launch attacks with minimum data perturbation to mislead the fusion center with a high probability.

Our idea of the surrogate model is partially based on \cite{papernot2017practical,shokri2017membership, Shi2017Surrogate}. Provided that the learning capacity of the surrogate model is equivalent to or stronger than the decision model at the fusion center, it will be easier to gradually learn an approximation of the decision model and is also practically feasible as it does not require the prior knowledge of the decision model or training data.

%

\subsection{LEB Attack Architecture}\label{architecture}
\begin{figure}[t]
\centering
\includegraphics[scale=1]{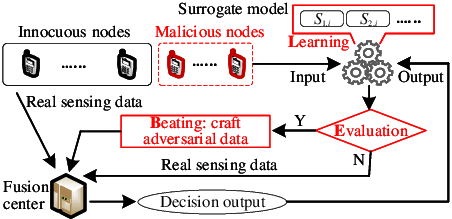}
\caption{The LEB attack framework.}
\label{Fig:framework}
\end{figure}

By leveraging the three observations in Section \ref{sec:attackmotivation}, we propose the \textit{Learning-Evaluation-Beating (LEB)} attack framework based on adversarial machine learning to launch attacks against cooperative spectrum sensing. As the name indicates, the LEB attack consists of three steps: (i) \textit{learning}, in which the attacker uses incremental learning models to build its own surrogate model to approximate the fusion center's decision model; (ii) \textit{evaluation}, in which the attacker evaluates whether the selected surrogate model is accurate enough to launch attacks, and (iii) \textit{beating}, in which adversarial sensing results are generated to fool the decision model. The flow chart of the LEB attack framework is shown in Fig.~\ref{Fig:framework}, \hlb{in which the red-colored components are controlled by the attacker. It is worth noting that the attacker is defined as a malicious controller that can access all the reports of controlled and compromised nodes, and modify the information that each device reports to the fusion center.}

\subsubsection{Learning}
The learning step aims to build the surrogate model $S_i$ at timeslot $i$ to approximate the decision model $O$ at the fusion center given malicious nodes' reporting vector $\{\mathbf{a}_j\}_{j\in[0, i]}$ and the fusion center's decision $\{y_j\}_{j\in[0, i]}$. As the LEB attacker has no initial knowledge of the decision model $O$, the learning step of the LEB attack should be generic and flexible. Our learning idea is inspired by \textit{no free lunch theorems} and \textit{transferability}.

The basic idea of all versions of \textit{no free lunch theorems} presented in  \cite{wolpert1997no} is that for both deterministic and stochastic  algorithms, ``the average performance of any pair of learning algorithms across all possible problems is identical", or more explicitly, it demonstrates that ``what an algorithm gains in performance on one class of problems is necessarily offset by its performance on the remaining problems".

No free lunch theorems imply that learning algorithms perform well on one problem does not mean it can always perform well when the problem changes, and that it is not a good idea to use a fixed learning model given different real-world scenarios. Therefore, we adopt a ``horses for courses" strategy rather than a ``once for all" strategy when choosing the surrogate model. We employ a set of learning algorithms as sub-models to comprehend varieties at the fusion center. At each timeslot, we choose the learning algorithm that has the best performance as the final surrogate model.

\textit{Transferability} shows that as long as models are trained to perform the same task, the influence of adversarial samples for one model can often be transferred to other models, even if they have different architectures or are trained on different training parameters or datasets \cite{papernot2016transferability}. Transferability property offers the justification of using adversarial samples generated from one sub-model to cheat other sub-models.

As a result, we are motivated to use a set of different machine learning models together to approximate the decision model $O$. Specifically, the surrogate model $S_i$ consists of $L$ machine learning models, called sub-models. Each sub-model is denoted as
\begin{equation}
S_{l,i}:\mathcal{A} \rightarrow \mathcal{Y}, \mathcal{A} \subset \mathbb{R}^{m \times 1}, \mathcal{Y} = \{-1,1\}, l = 1,2, \cdots, L,
\end{equation}
where $m$ is the number of malicious nodes controlled by the attacker. The number of sub-models $L$ depends on the attacker's resource and strength. When $L$ is large, the attacker will have more options in choosing the best surrogate model. Each sub-model is a particular representative machine learning model (e.g., Support Vector Machine (SVM), linear regression, multi-layer neural network (MLP), etc.) and is trained on the same set $\{<\mathbf{a}_j, y_j>_{j\in[0, i]}$\}.

\subsubsection{Evaluation}
As shown in Fig.~\ref{Fig:framework}, even we use all of the malicious nodes together to build a surrogate model, the model is only a partial one because the fusion center will also use sensing data from other innocuous nodes, which is unknown to the attacker. Thus, the goal of the evaluation procedure is to (i) evaluate whether the partial model (sub-models in the surrogate model) is accurate enough and (ii) use model selection to select the best sub-model to launch the attack.

At timeslot $i$, the attacker uses each of the sub-model $S_{l,i}$ to classify the current data vector $\mathbf a_i$ and obtains its local decision $z_i = S_{l,i}(\mathbf a_i).$ Then the attacker compares $z_i$ with the fusion center's decision $y_i$, and maintains a metric called internal accuracy for each sub-model $S_{l,i}$,  which is defined as $A_{l,i} = \frac{1}{i}\sum_{j=0}^{i} \mathbf 1_{\{z_j = y_j\}} \in [0,1], $ where $\mathbf 1_{\{z_j = y_j\}}$ denotes the indicator function that has value of 1, if $z_j = y_j$, and value of 0 otherwise. The internal accuracy measures how accurate a sub-model can track the fusion center's decision model.

At the evaluation step, if the highest internal accuracy in all sub-models is greater than a given threshold $\alpha$, the attacker will select the sub-model that has the highest internal accuracy as its final surrogate model for timeslot~$i$ and enters the beating procedure for the $i$th timeslot. Otherwise, the attacker will not attack, but simply send the real sensing data to the fusion center. This evaluation procedure ensures that the attacker will not attack with low confidence, which is also designed to improve the attack success probability under trust-based defenses.

We denote by $S_{l^*,i}$ the best sub-model with the highest internal accuracy selected at timeslot $i$, where $l^* = \arg\max_{l \in [1, L]} ~ A_{l,i} $. Then we formulate the surrogate model as

\begin{equation}
S_i = S_{l^*,i} = S_{\arg\max_{l \in [1, L]}  A_{l,i},i}.
\end{equation}

\subsubsection{Beating}
\hlb{In the learning and evaluation steps, the attacker passively listens and builds its model, and does not enter the beating step and launch any active attack unless it passes the threshold test in the evaluation step.} The goal of the beating step is for the attacker to craft adversarial sensing results based on the selected sub-model to beat the decision model at the fusion center and get a desired output. Attacker's adversarial sensing results are denoted as $\mathbf{a}^*_i$. We write $\mathbf{a}^*_i = \mathbf{a}_i + \bm\delta_i$, where $\mathbf{a}_i$ is the real sensing data and $\bm\delta_i$ is the adversarial perturbation.

\hlb{The beating step finds the best $\bm\delta_i$ with the minimum data perturbation satisfying
\begin{equation}\label{Eq:Opt}
\begin{aligned}
  &  \text{Objective:} & & \underset{\bm\delta_i\in \mathcal D}{\arg\min} \| \bm\delta_i \|_2, \\
  &  \text{Subject to:} & ~~~~ & S_{l^*,i}(\mathbf{a}^*
  _i) = S_{l^*,i}(\mathbf a_i + \bm \delta_i) \neq S_{l^*,i}(\mathbf{a}_i),
\end{aligned}
\end{equation}
where $\|\cdot\|_2$ is the L2-norm and $\mathcal D$ is the feasible solution space of $\bm\delta_i$, which is a constraint put onto $\textbf{a}^*$ to limit the maximum report $\textbf{a}_{\text{max}}$ and minimum report $\textbf{a}_{\text{min}}$,  which could alleviate the risk of being detected by defenses like outlier-based malicious detection \cite{kaligineedi2010malicious}. As the objective is to minimize the perturbation of given inputs, we found from real-world experiments that it can also alleviate much of other Byzantine detection methods \cite{li2010catch,penna2012detecting}.}

We note that the above optimization objective does not always guarantee a solution as a result of restriction from feasible solution space $\mathcal D$. More information about how to generate $\bm\delta$ under restriction $\mathcal D$ is detailed in Section \ref{SubSec:ADGen}.



After the beating process, all sub-models in the LEB attack framework need to be updated continuously over time and will be retrained by adding new input and output data in each round of spectrum sensing, which is the new learning step of the next episode. The nature of such a retraining process is to incrementally add training data to its already trained model, such that a more accurate model can be found over time. Therefore, the LEB attack is designed to employ online/incremental sub-models to efficiently update the surrogate model, such that it can be updated with new sensing data in a dynamic network environment.



\subsection{Generic Adversarial Sensing Data Generation}\label{SubSec:ADGen}
As discussed in the beating step, we must find the best $\bm\delta_i$ with the minimum data perturbation to satisfy Eq. \eqref{Eq:Opt}. When the best sub-model is chosen as the surrogate model $S_i = S_{l^*,i}$, $\bm\delta_i$ can be found via solving a method-specific optimization problem. For example, we can use the fast gradient sign methods (FGSM) \cite{goodfellow2014explaining} or the Jacobian-based saliency map approach (JBSM) \cite{papernot2016limitations} to find the best $\bm\delta_i$ if the selected sub-model is a deep neural network. However, this complicates the solution to Eq. \eqref{Eq:Opt} in the beating step, makes it dependent on a specific type of sub-model, and makes it less flexible and generic as the surrogate model contains $L$ sub-models, which is extendable by design in the LEB attack framework.

We aim to provide a generic adversarial sensing data generation algorithm to solve Eq. \eqref{Eq:Opt}. Our design intuition has two main components: (i) Unlike complicated data representation (e.g., image and voice data), signal strength data provides straightforward information: a larger value more likely indicates that a channel is occupied and a smaller value indicates otherwise. Therefore, searching is biased towards one direction in Eq. \eqref{Eq:Opt}. (ii) Transferability in machine learning indicates that if we can find adversarial examples of one sub-model, we can transfer it to other sub-models.

\begin{figure}
\centering
\includegraphics[scale=0.48]{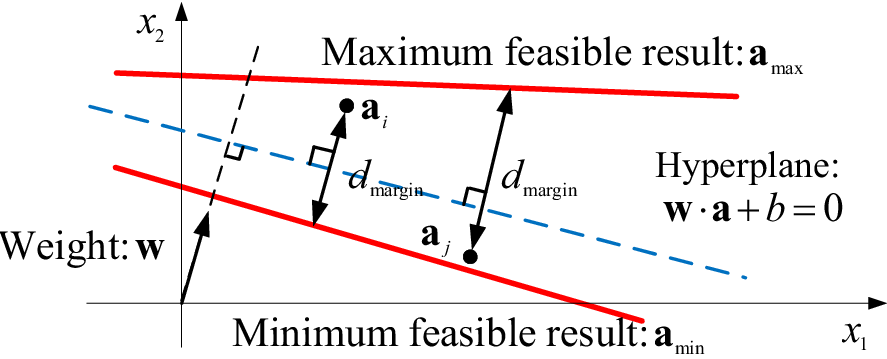}
\caption{Pilot model-based adversarial sensing result generation.}
\label{Fig:svm}
\end{figure}

However, in the LEB attack framework, different final surrogate models will be selected over time. Inspired from that, we propose a pilot model-based method for adversarial sensing result generation. This method is specifically designed for the cooperative spectrum sensing scenario, which consists of two steps: \textit{Step 1,} estimating the decision hyper-plane of training sensing results $\{<\textbf{a}_j, y_j>_{j\in [0,i]} \}$ from malicious nodes, which is determined by $\textbf{w} \cdot \textbf{a} + b = 0$ for a small set of sensing results ($\textbf{a}$ comes from $\{\textbf{a}_j\}_{j\in [0,i]}, $ also known as support vectors in SVM \cite{cortes1995support}) from the training data of the surrogate model. $\textbf{w}$ is the trained weight vector and $b$ is the bias (or intercept) (as shown in Fig.~\ref{Fig:svm}, in which we consider a two-dimension situation, denoted as $x_1$ and $x_2$); \textit{Step~2}, using a binary search structure along the direction defined by $\textbf{w}$ to find the final $\bm\delta_i$ to form the adversarial report $\mathbf{a}^*$.

\begin{algorithm}[t]
\DontPrintSemicolon
\SetAlgoLined
\SetKwInOut{Input}{Input}\SetKwInOut{Output}{Output}
\Input{\text{Sensing result $\textbf{a}_i$}, \text{feasible solution space} $\mathcal D$, \text{selected sub-model} $S_i$, algorithm termination threshold $\epsilon$, trained pilot model parameters: $\textbf{w},b$;}
\Output{\text{Adversarial perturbation} $\bm\delta_i$;}
 \BlankLine
Set two boundary vectors $\textbf{a}_{\text{min}}$ and $\textbf{a}_{\text{max}}$ according to $\mathcal D$; \\
Set $sgn = -sign(\textbf{w} \cdot \textbf{a}_i + b)$;\\
Set initial value: $\bm \delta_{i} \leftarrow sgn \times d_{\text{margin}}  \textbf{w}$;  \\

{\bf if} $S_i(\mathbf a_i + \bm \delta_i) \neq S_i(\mathbf{a}_i)$, {\bf then} \\
~~~~ $ l \leftarrow 0$; $r \leftarrow d_{\text{margin}}$; \\
~~~~{\bf repeat}\\
~~~~ ~~~~ {\bf if} $S_i(\mathbf a_i + (sgn (l + r)/2) \textbf{w}) \neq S_i(\mathbf{a}_i)$, {\bf then} \\
~~~~ ~~~~ ~~~~ $ r \leftarrow l + (r - l)/2$; \\
~~~~ ~~~~ {\bf else} \\
~~~~ ~~~~ ~~~~ $l \leftarrow l + (r - l)/2$; \\
~~~~ ~~~~{\bf end if}\\
~~~~ {\bf until} $\|r - l\| \leq \epsilon$;\\
~~~~ {\bf if} $S_i(\mathbf a_i + sgn \times r \textbf{w}) \neq S_i(\mathbf{a}_i)$, {\bf then} \\
~~~~ ~~~~ Return the adversarial perturbation: \\
~~~~ ~~~~ $\bm\delta_i \leftarrow sgn \times r \textbf{w}$. \\
~~~~ {\bf else} \\
~~~~ ~~~~ {terminate without any feasible solution.} \\
~~~~{\bf end if}\\
{\bf else} \\
~~~~ {terminate without any feasible solution.} \\
{\bf end if}
\caption{Adversarial Sensing Result Generation}
\label{adv_gen}
\end{algorithm}

The procedure of the proposed method is shown in Algorithm~\ref{adv_gen}. We can use SVM as the pilot model due to its strong transferability performance \cite{papernot2016transferability} in practice. However, it is not necessary to train an extra pilot model; instead, we can choose the sub-model that has the best consistency in terms of internal accuracy with the fusion center as the pilot model among models that involve learning a hyper-plane (such as Passive Aggressive algorithm \cite{crammer2006online} and other classifiers). Because of the logarithmic complexity ($O(log N)$ with searching size $N$) of the binary search structure, the proposed algorithm is expected to yield a fast and efficient solution to Eq. \eqref{Eq:Opt} in cooperative spectrum sensing scenarios.

\section{Experiment Evaluation} \label{exper_eval}

In this section, we present our real-world dataset collected from practical TV white space signal strengths. Based on this dataset, we conducted experiments to measure the impact of LEB attacks under various conditions.

\subsection{Experimental Setup} \label{sec:experiment}
We present measurement details, configurations of the fusion center, the LEB attack framework, and performance metrics in this subsection.
\subsubsection{Measurement configurations}
We collected realistic TV white space signal strengths using RTL-SDR TV dongles, which have been validated to have adequate signal detection capabilities \cite{saeed2017local}. We deployed 20 RTL-SDR TV dongles as sensing nodes on a campus area to collect signal strengths simultaneously on TV channels based on configurations used in \cite{saeed2017local}. We used GNURadio v3.7.9 to implement the sensing process for each dongle that uses the averaged signal power over a time period of 30 seconds (required by Federal Communications Commission \cite{FCC2015}) as the sensing result for one timeslot.

\begin{figure}
\centering
\includegraphics[scale=1]{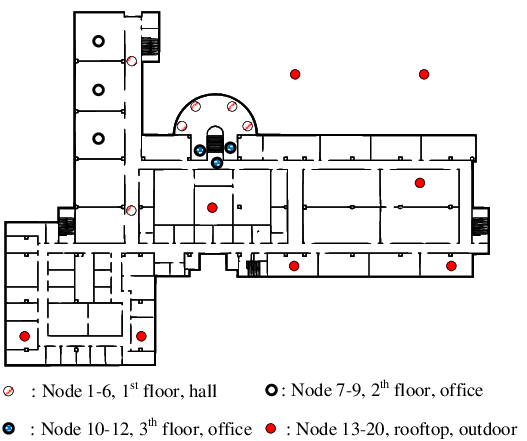}
\caption{The deployment environment of 20 RTL-SDRs.}
\label{Location}
\end{figure}

To make the collected signal data comprehend the realistic spectrum sensing scenarios as much as possible, we deployed sensing devices in various surrounding environments. Fig.~\ref{Location} shows how 20 RTL-SDR TV dongles were deployed throughout a $379 \times 232$ $\text{ft}^2$ building: 8 TV dongles were placed outside of the building and 12 were distributed within the building on different floors and indoor environments. We distributed TV dongles at these various places and environments to represent different spectrum sensing scenarios in practice, which is similar with the deployment style of CORNET testbed \cite{TestBed}. We collected signal strengths on 22 different channels for 100 continuous hours on each sensing node, such that the dynamic model mismatch phenomenon can be recorded in the dataset.

\subsubsection{Fusion center configurations}
To show the attack performance of the LEB attack framework, we implemented eight existing representative intrusion detection defenses at the fusion center. Specifically, four statistics-based defenses: Outlier factors based defense (Outlier) \cite{kaligineedi2010malicious}, Local Outlier Factor based defense (LOF) \cite{amini2014local}, Empirical Covariance based (EmpCov) and Robust Covariance based (RobCov)) detections \cite{rousseeuw1999fast}; three machine learning based defenses: fuzzy kNN based defense (fzKNN) \cite{wang2018primary}, Double-Sided Neighbor Distance based defense (DSND) \cite{li2010catch} and One-class SVM based detection (OCSVM) \cite{rajasegarar2015pattern}; and one trust-based detection method (Trust) \cite{chen2008robust}.



\hlb{There are various data fusion rules such as SVM, Logistic Regression (LR), AND, OR and Majority rule\cite{jiang2013joint,amini2014local,fatemieh2011using, saeedlocal}. Due to the LEB attack design and the property of transferability in machine learning, the fusion rule does not generally affect the design of the LEB attack architecture, but may affect its attack result. For example, the majority rule may perform better than SVM against the LEB attack (as the attack is likely to fail when malicious nodes are not the majority) but it does not generate good performance as pointed out in the literature \cite{fatemieh2011using, wang2018trust}. We choose the SVM rule as the representative rule in our experimental evaluation because (i) it is one of the most widely used fusion rules for efficiency and performance \cite{fatemieh2011using, saeedlocal}, and (ii) our main goal is to show that how a machine learning powered attacker can take advantage of the open nature of cooperative dynamic spectrum sensing to launch a new type of attack.}

We used the sensed data of the first five hours collected from 22 nodes and the corresponding ground truths of the TV channel status in the local area to build statistical models and as training data for the defense methods implemented at the fusion center. The remaining data was used as test data.


\begin{figure*}
\centering
\includegraphics[scale=1]{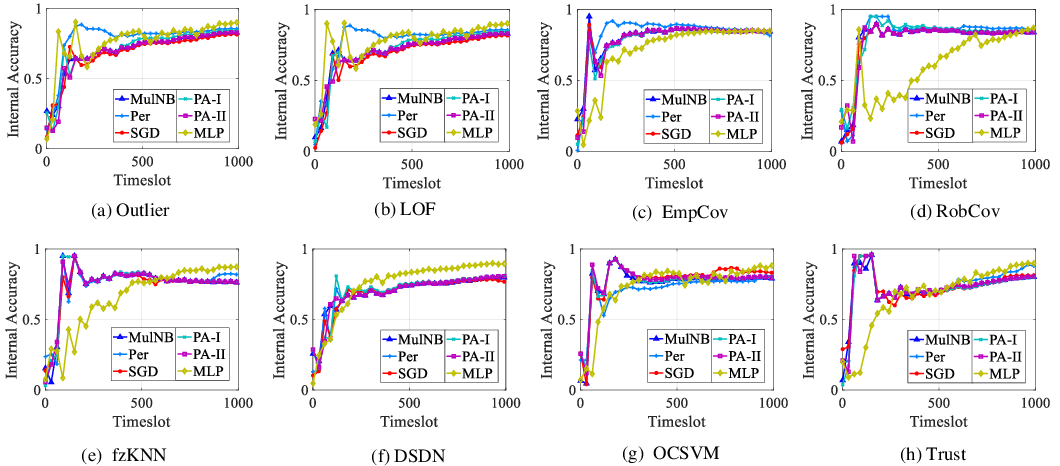}
\caption{Internal accuracy transitions of the LEB attacker under different defenses.}
\label{defenses_acc}
\end{figure*}

\subsubsection{LEB attack framework configurations}

We implemented the surrogate model based on manipulated nodes, which consists of six incremental learning based sub-models: Naive Bayes classifier for multinomial models (MulNB), Perceptron classifier (Per), Linear SVM classifier with stochastic gradient descent training (SGD), Passive Aggressive-I classifier (PA-I), Passive Aggressive-II classifier (AP-II) and Multi-layer Perceptron classifier (MLP). These six models are well-known and representative techniques in incremental learning. In practice, we can add other incremental learning-based algorithms or exclude models based on security requirements of specific scenarios.

Without loss of generality, we let the interaction of cooperative spectrum sensing and the LEB attacker that controls the malicious nodes start at timeslot 0 corresponding to the $6$th hour of the collected data. We assume that the defense mechanism has already been trained using the first five hours of the collected data.  At each timeslot that takes 30 seconds, the LEB attacker learns and evaluates the surrogate models to launch a potential attack.

In the evaluation step, the internal accuracy threshold $\alpha$ is 0.85 unless otherwise specified. In the beating step, the LEB attacker generates adversarial sensing data based on the generic generation algorithm proposed in Section~\ref{SubSec:ADGen}.

\subsubsection{Performance metrics}
We evaluated the performance of the LEB attacker through two metrics: \textit{attack success ratio} and \textit{overall disruption ratio}. We define the attack success ratio as the ratio of the number of attacks that successfully mislead the fusion center to make wrong decisions to the number of attack attempts. We define the overall disruption ratio as the ratio of the number of successful attacks to the number of elapsed timeslots, i.e.
\begin{equation} \label{eq:def}
\begin{split}
&\text{attack success ratio} = \frac{\# \text{of successful attacks}}{\# \text{of attack attempts}}, \\
&\text{overall disruption ratio} = \frac{\# \text{of successful attacks}}{\# \text{of elapsed timeslots}}.
\end{split}
\end{equation}
Note that a higher attack success ratio does not necessarily mean a higher overall disruption ratio. The reason is that when the LEB attacker does not pass the evaluation step (e.g., it may have a large threshold $\alpha$ to make an attack attempt succeed with high probability), it will not launch the attack and thus will not cause a disruption to the network. In comparison, the attack success ratio measures the learning and evaluation quality of the LEB attack framework, and the overall disruption ratio quantifies the performance impact that the LEB attacker brings to the entire cooperative sensing network.

\subsection{Results and Analysis}
We show experimental results of the LEB attack in terms of attack success ratio and overall disruption ratio measurements in this subsection.
\subsubsection{Attack impacts on defense strategies}
We first evaluated the impact of the LEB attacker on each of the 8 defense strategies used by the fusion center. We randomly selected 8 nodes as malicious nodes to be controlled by the LEB attacker. The LEB attacker updates its internal accuracies for each sub-model to evaluate and beat the fusion center. \hlb{As we do not assume any specific statistical/stationary model for the behavior of the attacker and nodes, sensing reports, or wireless spectrum properties, it can be mathematically infeasible to accurately formulate the interactions between the attacker and the fusion center or perform analytical convergence analysis. Our experimental evaluation results in Fig.~\ref{defenses_acc} shows the attacker's internal accuracies and the convergence trends of sub-models with regard to the timeslot against defense strategies (a) Outlier, (b) LOF, (c) EmpCov, (d) RobCov, (e) fzKNN, (f) DSDN, (g) OCSVM and (h) Trust.}

We observe from Fig.~\ref{defenses_acc} that when the attacker starts to learn, internal accuracies of each sub-model change drastically; but with more timeslots elapsed, they gradually become stable. For example, in Fig.~\ref{defenses_acc} (c), accuracies start to stabilize at around 0.85 after 500 timeslots, which is the approximated convergence time in the training process of the surrogate model.

\begin{figure}[t]
\centering
\includegraphics[scale = 0.47]{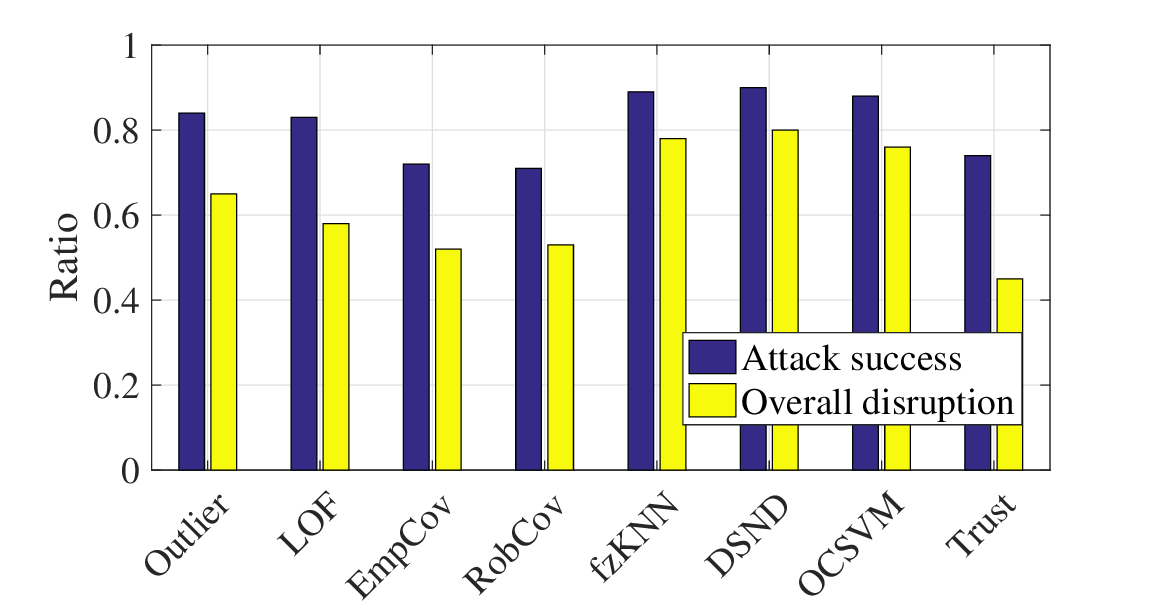}
\caption{The Attack success and overall disruption ratios of the LEB attacker with different defense strategies.}
\label{overall_def}
\end{figure}

Fig.~\ref{overall_def} illustrates the attack success ratio and overall disruption ratio of the LEB attacker against each defense method individually. From Fig.~\ref{overall_def}, we note that the LEB attacker achieves 71\%--90\% attack success ratios against different methods, which means that the learning and evaluation design in the LEB attack framework is effective for the attacker to assess its potential capability to launch successful attacks. We also observe from Fig.~\ref{overall_def} that the overall disruption ratio is 45\%--80\% due to the attack, which indicates that the LEB attacker is able to successfully fool the fusion center into making wrong decisions for 45\%--80\% of the time, thereby resulting in severe performance disruption.

\subsubsection{Impacts of threshold $\alpha$}
A key factor in the LEB attack framework is the internal accuracy threshold $\alpha$. The attacker can only launch attacks if the internal accuracy of a sub-model exceeds $\alpha$. A larger $\alpha$ should lead to a higher attack success ratio but should also decrease the overall disruption ratio, which is because a larger $\alpha$ will decrease attack attempts but result in a better estimation about the decision rule.

\begin{figure}[t]
\centering
\includegraphics[scale=0.85]{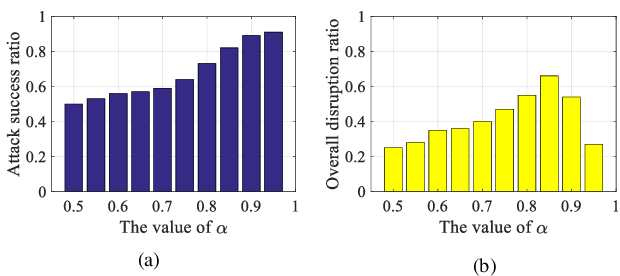}
\caption{The relationship between $\alpha$ and the attack success ratio of the LEB attacker.}
\label{Fig:threshold}
\end{figure}

Fig.~\ref{Fig:threshold} depicts (a) the attack success ratio and (b) the overall disruption ratio for different values of $\alpha$ (the rest of the setting is the same as that in Fig.~\ref{overall_def} and results are averaged over eight defense strategies).  We observe from Fig.~\ref{Fig:threshold}(a) that when $\alpha$ is 0.7 or less, the attack success ratio is around 0.5. As $\alpha$ increases, the attack success ratio also increases. The reason is that an increased $\alpha$ requirement indicates a higher consistency or better estimation of the fusion center for the selected sub-model.

However, in terms of the overall disruption ratio, although a smaller $\alpha$ will lead to more attack attempts, it will also increase the risk for the attacker to activate defense mechanisms at the fusion center, further decreasing the overall disruption ratio. The overall disruption ratio increases when $\alpha$ is small and then decreases as the threshold $\alpha$ approaches to 1, as shown in Fig.~\ref{Fig:threshold}(b). The reason is that as the threshold $\alpha$ approaches to 1, the number of attack attempts decreases significantly. Thus, although the attack success ratio increases, the overall number of attacks decreases. We observe from our experiments that the optimal threshold can be achieved at around $\alpha=0.85$ to maximize the overall disruption ratio.

\subsubsection{Impacts of the number of malicious nodes}
The number of malicious nodes controlled by the LEB attacker is a fundamental factor that affects the attack impact, which answers the question of how many nodes an attacker needs to achieve a promising attack performance. Table~\ref{Tab:node_scale} shows the overall disruption ratio against different defenses when the number of randomly selected malicious nodes increases from 2 to 10 (i.e., 10\% to 50\% of all nodes). Table~\ref{Tab:node_scale} shows that as the number of malicious nodes approaches 10, which is half of the total sensing nodes, the overall disruption ratio (averaged over all defenses) reaches 72\%, which means that the spectrum sensing is disrupted by the attacker at 72\% of the elapsed timeslots. The attack impact can still be observed even when the number of malicious nodes is small. For example, three malicious nodes (15\% of all nodes) can lead to a nearly 20\% overall disruption ratio.

\begin{table}[t]
\begin{center}
\caption{Overall disruption ratios (\%) under different scenarios.}\label{Tab:node_scale}
\begin{tabular}{c|ccccccccc}
\hline
\multirow{2}{*}{\textbf{Defenses}} &\multicolumn{9}{c}{\textbf{The number of malicious nodes $m$}}   \\ \cline{2-10}  & \textbf{2} & \textbf{3} & \textbf{4}  &  \textbf{5} & \textbf{6} &  \textbf{7} & \textbf{8} & \textbf{9} & \textbf{10}\\\hline
Outlier    & 8 & 20 & 23 & 24 & 34 & 38 & 65 & 68 & 72\\
\hline
LOF        & 6 & 15 & 23 & 25 & 35 & 36 & 58 & 60 & 65\\
\hline
EmpCov     & 9 & 22 & 27 & 35 & 37 & 39 & 52 & 59 & 66\\
\hline
RobCov     & 7 & 15 & 22 & 30 & 39 & 41 & 53 & 56 & 64\\
\hline
fzKNN     & 10 & 21 & 27 & 45 & 45 & 49 & 78 & 80 & 82\\
\hline
DSND      & 11 & 23 & 31 & 44 & 46 & 47 & 80 & 85 & 86\\
\hline
OCSVM      & 9 & 21 & 31 & 39 & 48 & 51 & 76 & 79 & 80\\
\hline
Trust      & 6 & 16 & 21 & 29 & 34 & 38 & 45 & 56 & 60\\
\hline
\textbf{Average} & 8 & 19 & 28 & 34 & 40 & 42 & 63 & 69 & 72\\
\hline
\end{tabular}
\end{center}
\end{table}

It can be concluded from Table~\ref{Tab:node_scale} that the LEB attack framework provides an effective attack strategy even when the number of malicious nodes is small. The attack performance also improves as the number of malicious nodes increases.

\subsubsection{Impacts of locations of malicious nodes}
In previous experiments, we always randomly selected nodes as malicious ones. We are also interested in whether malicious nodes can bring more impact to the network if they choose to be at ``better'' locations. We divide the dongles into four groups with five in each group. The experiments are conducted by using one group as malicious nodes while the rest being innocuous.

\begin{figure}[t]
\centering
\includegraphics[scale=0.48]{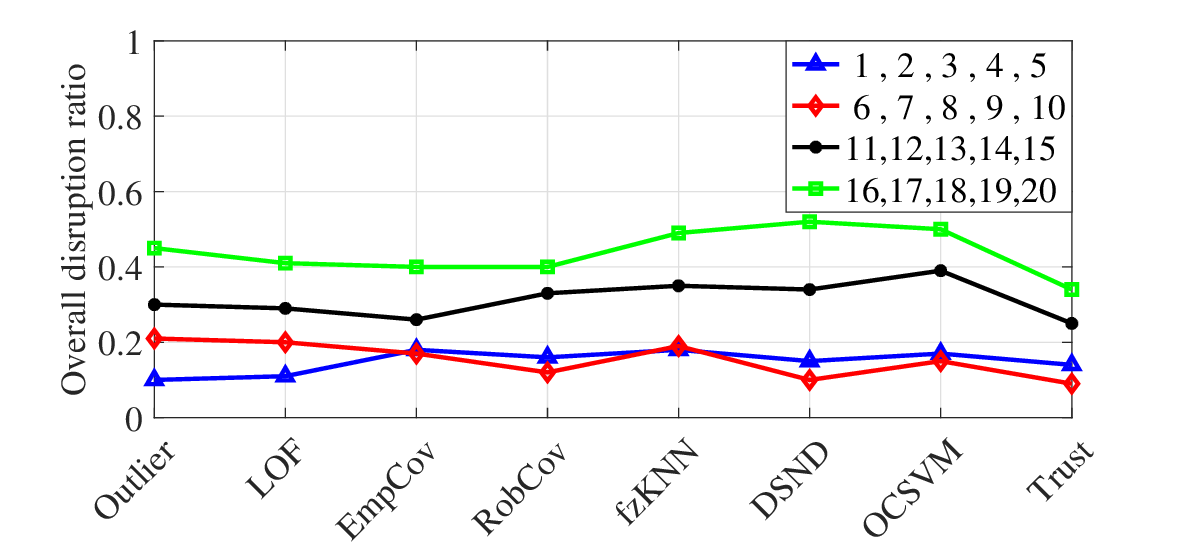}
\caption{Overall disruption ratios when manipulating different sets of nodes.}
\label{Fig:diff_nodes}
\end{figure}

The results are depicted in Fig.~\ref{Fig:diff_nodes}, from which we observe that when controlling nodes 16--20 (distributed in outside environment) as malicious ones, the overall disruption ratio is 43\% averaged over all eight defenses. However, if nodes 6--10 (distributed in indoor environment) are controlled as malicious ones, the averaged overall disruption ratio is only 17\%. Hence, we can conclude that it is critical for malicious nodes controlled by the LEB attacker to be at ``right'' locations that have a higher influence on the decision process in order to launch more effective attacks.

\subsubsection{Efficiency of adversarial sensing data generation}
We also compare the performance of the proposed method for adversarial sensing result generation with other adversarial sample generation methods: Fast gradient sign method (FGSM) \cite{goodfellow2014explaining}, Jacobian-based saliency map approach (JBSM) \cite{papernot2016limitations}, DeepFool method  (DF) \cite{moosavi2016deepfool},  Basic iterative method (BaIter) \cite{kurakin2016adversarial}, SPSA attack method (SPSA) \cite{uesato2018adversarial} and Elastic-Net method (EN) \cite{chen2017ead} in Deep Neural Network, which are implemented based on CleverHans V2.1.0 \cite{papernot2018cleverhans}.

\begin{figure}[t]
\centering
\includegraphics[scale=0.9]{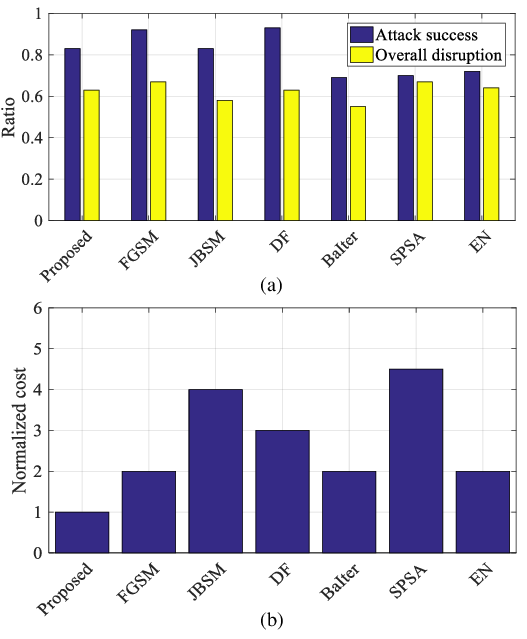}
\caption{Comparison of the adversarial sensing result generation methods.}
\label{Fig:adv_gen_methods}
\end{figure}

In these experiments, 8 nodes are randomly selected as malicious nodes and they use different adversarial data generation methods in the LEB attack framework to generate malicious data inputs to the fusion center. Fig.~\ref{Fig:adv_gen_methods} shows (a) the attack success ratio and overall disruption ratio under different generation methods and (b) the normalized costs of the generation methods (the normalized cost of a generation method is defined as its computational time to generate the adversarial data vector divided by the computational time of our proposed method to generate the adversarial data; thus our proposed method has a normalized cost of 1).

Fig.~\ref{Fig:adv_gen_methods}(a) shows that our proposed method has the similar attack success ratio and overall disruption ratio as other methods. Fig.~\ref{Fig:adv_gen_methods}(b) further shows that our method is much more computationally efficient than the other methods. For example, SPSA has a normalized cost of around 4.5, and the overall average computation cost reduction compared to other methods is around 65\%, which is mainly contributed by the simplicity and logarithmic complexity of binary search method in our proposed method.

\subsubsection{Performance comparison with existing attacks}
The LEB attack is a new type of attack against spectrum sensing. There are many other proposed Byzantine attacks against the fusion center. They can be roughly classified into two categories: Independent Malicious Byzantine Attacks (IMBA) and Cooperative Malicious Byzantine Attacks (CMBA) \cite{rawat2010collaborative, li2010catch,penna2011detecting}. In IMBA, each Byzantine node makes the decision to attack the fusion center independently, while Byzantine nodes collaborate with each other in CMBA.

IMBA is a comparatively simple attack strategy. It requires no cooperation between the nodes and each individual node makes the decision by itself. It is proved in \cite{rawat2010collaborative} that unless the number of Byzantines is larger than or equal to 50\% of the total nodes, the fusion center cannot be made blind by attackers. More information of the IMBA strategy can be found in \cite{li2010catch, penna2011detecting}.

In the CMBA scenario, Byzantine nodes cooperate with each other and make the decision collaboratively to attack the fusion center, which is a much stronger attack model than IMBA. Therefore, the number of malicious nodes required to make the fusion center blind is also less than 50\% \cite{rawat2010collaborative}. LEB attacks belong to the CMBA category, but the attack model is much weaker than the CMBA model in \cite{li2010catch}. The LEB attack requires no information of other normal nodes, including the total number of sensing nodes and the sensing results.

In this group of experiments, we compare attack performance of LEB attacks with both IMBA and CMBA methods. Rather than using simple data fusion methods like ``AND'' \cite{peh2007optimization}, ``OR'' \cite{ghasemi2005collaborative}, or ``Weighted'' \cite{quan2008optimal}, we employ a more comprehensive machine learning based method, namely SVM as the decision rule \cite{saeed2017local}.

Existing attack methods that we choose for comparison include an independent malicious attack model identified in \cite{li2010catch}, in which each malicious node launches the attack according to a preset probability; an adaptive probabilistic independent attack model that considers attack costs proposed in \cite{ahmadfard2017probabilistic}; and a cooperative dependent malicious attack model proposed in \cite{qin2013defending}, in which an attacker injects malicious sensing data simultaneously and the falsified data is self-consistent.

Attack success ratios and overall disruption ratios are used to quantify the attack performance. The detailed experimental results are shown in Table \ref{Tab:diff_atks}, in which 8 out of 20 nodes are malicious. We can observe that cooperative attacks (\cite{qin2013defending} and LEB attacks) generally have better performance than independent attacks (\cite{li2010catch,ahmadfard2017probabilistic}), on average more than 10\% of improved performance in terms of both attack success ratio and overall disruption ratio.

Note that unlike the trust-based fusion, the fusion center is trained using SVM. LEB attacks cannot change the values of parameters in the decision model. LEB attacks still have slightly better overall disruption ratio than \cite{qin2013defending} but they have better attack success ratio (around 10\% of the advantage) due to the learning characteristics of tLEB attacks.

\begin{table}[t]
\begin{center}
\centering
\caption{Attack success ratios (\%) and overall disruption ratios (\%) comparison under different attack methods.}\label{Tab:diff_atks}
\begin{tabular}{c|c|c}
\hline
\textbf{Attack methods} & \textbf{Attack success} & \textbf{Overall disruption}  \\
\hline
Li et al.\cite{li2010catch}    & 76 & 76  \\
\hline
Ahmadfard et al. \cite{ahmadfard2017probabilistic}   & 73 & 73   \\
\hline
Qin et al. \cite{qin2013defending}   & 84 &  84 \\
\hline
LEB attacks    & 95 &  85 \\
\hline
\end{tabular}
\end{center}
\end{table}

\textbf{Summary: } We conducted comprehensive experiments based on real-world collected spectrum data to validate the LEB attack framework and compared the attack performance with existing defenses and attacks. The principle benefit of the LEB attack framework is due to learning, which supports the attacker to adapt and take advantage of the model mismatch phenomenon in practice.

\section{Influence-Limiting Defense} \label{sec:defense}
\hlb{The LEB attack poses a new security threat to cooperative spectrum sensing. In this section, we propose a countermeasure policy named influence-limiting defense to combat LEB attacks. We design the influence-limiting defense as a non-invasive policy such that it can coexist with traditional defenses \cite{li2010catch,chen2008robust,kaligineedi2010malicious,fatemieh2011using,penna2012detecting}.
}
\subsection{Defense Motivation}
Many traditional defenses such as statistics-based defenses \cite{kaligineedi2010malicious,chen2017cooperative,penna2012detecting}, machine learning-based defenses \cite{fatemieh2011using,wang2018primary, li2010catch, rajasegarar2015pattern} and trust value-based methods \cite{rawat2011collaborative,chen2008robust, sagduyu2014trust} have been designed by making assumptions about attackers. In a scenario where learning-empowered attackers are present, there exist following concerns:
\begin{itemize}
\item  The identified model mismatch phenomenon that results from spatial and temporal unevenness will lead to dynamic changes of the statistical property at each sensing node.
\item  When the attacker decides not to launch the attack, it can make controlled nodes behave ``normal'' by correcting random mistakes and combining all the sensing results of manipulated nodes to make the channel status decision, which improves the statistical consistency of those controlled nodes with the fusion center at timeslots without attacks, such that they can maintain a competitive attack budget at timeslots with attacks.
\item When the manipulated nodes decide to launch the attack, they can learn an efficient way through the LEB attack framework, which minimizes the pattern deviations.
\end{itemize}

Although many traditional methods have been proposed to counter uncertainties of sensing results from each node (e.g., metrics of ``weight'', ``trust value'', and ``reputation'' are widely used to balance the individual decision of each node and the fusion center decision \cite{varshney1986optimal,zhang2015byzantine, li2010catch, chen2008robust, kaligineedi2010malicious, rawat2011collaborative, fatemieh2011using, wang2014secure, yan2012vulnerability}), most of these metrics are directly or indirectly based on the statistical consistencies of individual decisions with global decisions. Nevertheless, high consistencies do not guarantee high worthiness of trust. As we have detailed in the LEB attack strategy, malicious nodes can be controlled to purposely maintaining a high consistency with the fusion center to avoid being detected. 

To our best knowledge, robust and generic strategies to combat these challenges in cooperative spectrum sensing are largely missing. We focus on finding the reason why a limited number of malicious nodes can succeed after learning the fusion center's model. Our key observation is that malicious nodes can take advantage of the model mismatch phenomenon to build up their dominant roles in the decision inference process, i.e., malicious nodes can accumulate their influence on the fusion center, thus the influence of normal nodes are comparatively decreased. From the defense perspective, although we may not know identities of attackers in the network, we can limit the influence of any subset of nodes on the fusion center's global decision.

Our objective is to design a robust and generic countermeasure named  influence-limiting defense specifically for the spectrum sensing domain. The method aims to bridge the gap between traditional defense methods and new challenges posed by LEB attacks by introducing the decision-flipping influence.

\subsection{Defense Framework}
We detail the influence-limiting defense framework in the following subsections.

\subsubsection{Defense objective}
\hlb{To measure the influence of any given subset of nodes $\mathcal X_{\text{sub}}$, we propose a new metric called \textit{decision-flipping influence}, denoted as $I(\mathcal X_{\text{sub}})$,
\begin{equation} \label{eq:6}
\begin{split}
& I(\mathcal X_{\text{sub}}) = \frac{
\vspace{5mm}
\# \text{ of } \bf{a}^* }{\# \text{ of } \bf{a}},\\
& \text{s.t. } O(\textbf{x}^*) \neq O(\bf{x}),
 \end{split}
\end{equation}
in which $\bf{a}^*$ is from $\mathcal X_{\text{sub}}$, and is part of $\bf{x}^*$. Intuitively, (\ref{eq:6}) indicates the probability of finding a malicious input $\textbf{x}^*$ that flips the decision output given $\textbf{x}$ by changing $\textbf{a}$. Apparently, the influence of all nodes is 1, i.e., $I(\mathcal X) = 1$. Unlike ``weight'', ``trust value''  or ``reputation'' based measures and the influence-based measure proposed in \cite{jiang2013renewal,jiang2013joint}, the decision-flipping influence $I(\mathcal X_{\text{sub}})$ is a direct probabilistic measure of the role $\mathcal X_{\text{sub}}$ played in the final decision process.}

Starting from the decision-flipping influence $I(\mathcal X_{\text{sub}})$, we can mitigate the severe impact of potential LEB attacks by enforcing an influence-limiting policy, in which the decision-flipping influence of any subset $\mathcal X_{\text{sub}} \subset \mathcal X$ should satisfy $I(\mathcal X_{\text{sub}}) \leq \delta(|\mathcal X_{\text{sub}}|)$, where $|\mathcal X_{\text{sub}}|$ denotes the number of nodes in $\mathcal X_{\text{sub}}$, and $\delta(|\mathcal X_{\text{sub}}|)$ is the threshold function of $|\mathcal X_{\text{sub}}|$, i.e., the influence-limiting policy is triggered only when $I(\mathcal X_{\text{sub}})$ goes beyond $\delta(|\mathcal X_{\text{sub}}|)$. For example, $\delta(1)$ denotes the threshold to limit the influence of each individual node.

\hlb{From the LEB attack framework, we know that as malicious nodes keep succeeding to flip the decision of the fusion center, their corresponding decision-flipping influence $I(\mathcal X_{\text{sub}})$  will also increase, thus influence-limiting policy will be triggered. We can write the influence-limiting policy as follows:}
\hlb{\begin{equation}\label{Eq:framework}
\begin{split}
\text{Objective:}~~~~~~~& \text{minimize } (y-\hat{y})^2, \\
\text{Subject to:}~~~~~~~& I(\mathcal X_{\text{sub}}) \leq \delta(|\mathcal X_{\text{sub}}|), ~~ \forall \mathcal X_{\text{sub}} \subset \mathcal X,
\end{split}
\end{equation}
in which $y$ is the true channel status and $\hat{y}$ denotes the predicted decision output. This is a non-invasive method that can coexist with traditional/existing defenses as mentioned in Section 2.2.}
\subsubsection{Finding threshold function $\delta(|\mathcal X_{\text{sub}}|)$}

The threshold function of $\delta(|\mathcal X_{\text{sub}}|)$ with regard to $|\mathcal X_{\text{sub}}|$ is the core component of the influence-limiting defense. To choose the threshold function, we first discuss simple cases, then extend them to generic cases:


\hlb{\textit{(i) In a well-balanced cooperative spectrum sensing scenario without malicious nodes}, the value of $\delta(|\mathcal X_{\text{sub}}|)$ in terms of $|\mathcal X_{\text{sub}}|$ intuitively needs to satisfy the following three basic requirements: (a) $\delta(|\mathcal X_{\text{sub}}|) \rightarrow 0$ when $|\mathcal X_{\text{sub}}| \rightarrow 0$; (b) $\delta(|\mathcal X_{\text{sub}}|)$ is monotonically increasing with regard to $|\mathcal X_{\text{sub}}|$; (c) $I(\mathcal X_{\text{sub}})  \rightarrow \frac{1}{2}$ when $|\mathcal X_{\text{sub}}| \rightarrow \frac{n}{2}$, which is to ensure that the influence over the fusion center is dominated by the majority rather than a small group of sensing nodes.}

\hlb{Based on the requirements mentioned above, a sigmoid style function comes out to be a feasible choice to interpolate the function of $\delta(|\mathcal X_{\text{sub}}|)$ with regard to $|\mathcal X_{\text{sub}}|$,
\begin{equation}
\delta(|\mathcal X_{\text{sub}}|)=\frac{1}{1+e^{-c_1(|\mathcal X_{\text{sub}}|-\frac{n}{2})}},  0 \leq |\mathcal X_{\text{sub}}| \leq \frac{n}{2},
\end{equation}
where $c_1$ is the control parameter used to adjust the function to better interpolate various practical scenarios. We assume that the number of malicious nodes is less than the number of normal nodes, which is the typical attack model \cite{zhang2015byzantine, li2010catch, kaligineedi2010malicious, wang2014secure, yan2012vulnerability}.}

\textit{(ii) In a generic scenario where malicious nodes may present}, the function $\delta(|\mathcal X_{\text{sub}}|)$ defined in (i) has no mechanism to counter malicious nodes. It is intuitive that when malicious nodes are present in $\mathcal X_{\text{sub}}$, $\delta(|\mathcal X_{\text{sub}}|)$ should be limited to a smaller threshold value such that the decision-flipping influence can be restrained. The next question is how one can know which node is malicious.

Note that due to the temporal or spatial unevenness, it is difficult to accurately identify which node is indeed malicious by checking the statistical property of the signal strengths. Therefore, instead of offering a hard decision rule to clearly classify a node into either innocuous or malicious, we design a soft rule to discriminate certain nodes in the final decision by the fusion center.

In particular, we still leverage a node's signal strength, but only checking changes of its statistical property. Suppose when a node's signal strengths exhibit different properties during the training and testing (or decision) phases, there exist following indications:
\begin{itemize}
\item The node may be malicious and its signal strengths are manipulated for an effective attack. If this is the case, the node should be at least less weighted (if not eliminated) in the fusion center's decision.
\item The node may be legitimate but its signal property changes due to the temporal or spatial unevenness, which further means that the original training data for this node does not reflect its current signal property and thus becomes less useful for the current decision.
\end{itemize}

In both cases, we should at least  weight those nodes less in the final decision. Therefore, we adopt the Kolmogorov-Smirnov (K-S) statistic to quantify such a change for a node. For node~$j$, the K-S statistic $d_{\text{ks}}^j$ is
\begin{equation} \label{eq:ks}
d^{j}_{\text{ks}} = \underset{x}{\text{sup}}|F_T^j(x)-F_{C}^j(x)|,
\end{equation}
where $F_T^j(x)$ is the training data distribution for node~$j$,  $F_{C}^j(x)$ is the empirical distribution of node~$j$'s signal strength in the test period, representing its current signal property, and $\text{sup}(\cdot)$ denotes the maximum value or upper bound.

$F_T^j(x)$ and $F_{C}^j(x)$ can be estimated from the sensed data. We can employ kernel density estimation method to obtain the distribution due to its flexibility in choosing kernel functions.  Based on Eq. \eqref{eq:ks}, we present our generalized influence-limiting policy as
\begin{equation}\label{eq:threshold}
\begin{split}
\delta(|\mathcal X_{\text{sub}}&|)=\frac{1}{1+e^{-c_1(|\mathcal X_{\text{sub}}|-\frac{n}{2})}}-c_2{\sum_{j \in \mathcal X_{\text{sub}}}d^{j}_\text{ks}}, \\
&0 \leq |\mathcal X_{\text{sub}}| \leq \frac{n}{2},   0 < \delta(|\mathcal X_{\text{sub}}|) < 1,
\end{split}
\end{equation}
where $c_1$ is a cost control parameter and $c_2$ is an influence control parameter.

The policy ensures that if there are more abnormal nodes in $\mathcal X_{\text{sub}}$, then  $\sum_{j \in \mathcal X_{\text{sub}}}d^{j}_\text{ks}$ will be larger and thus the threshold $\delta(|\mathcal X_{\text{sub}}|)$ will be smaller. When all the nodes in $\mathcal X_{\text{sub}}$ are normal nodes, $\sum_{j \in \mathcal X_{\text{sub}}}d^{j}_\text{ks}$ is expected to be small.

\subsubsection{Enforcing influence-limiting policy}
At the fusion center, the objective of the defense is to mitigate the impact posed by potentially malicious nodes through enforcing the influence-limiting policy, i.e., limiting the decision-flipping influence of any given $\mathcal X_{\text{sub}}$ within the range of $[0,\delta(|\mathcal X_{\text{sub}}|)]$.

\hlb{There are multiple ways to enforce the policy. One way is to re-weight the sensing results of $\mathcal X_{\text{sub}}$. Specifically, when the measured decision-flipping influence $I(\mathcal X_{\text{sub}})$ goes beyond $\delta(|\mathcal X_{\text{sub}}|)$ in a traditional defense method, the influence-limiting policy will limit the weight of $\mathcal X_{\text{sub}}$ to the threshold. In other words, the weight $w(\mathcal X_{\text{sub}})$ can be written as
\begin{equation}\label{eq:weight}
\begin{split}
w(\mathcal{X}_{\text{sub}}) = \left\{ \begin{array}{cc}
                \delta(|\mathcal X_{\text{sub}}|), & \hspace{1mm} I(\mathcal X_{\text{sub}}) \geq \delta(|\mathcal X_{\text{sub}}|), \\
                \text{Original weight,} & \hspace{1mm} \text{otherwise}.\\
                \end{array} \right.
\end{split}
\end{equation}
Another way to enforce this policy is to work with an existing defense such as the incentive method proposed in \cite{gan2016secure} to re-weight the corresponding nodes, thus the suspicious nodes can be discouraged and suppressed.}

\subsubsection{Balancing effectiveness and complexity}

It is worth noting that in order to achieve the objective defined in Eq. \eqref{Eq:framework}, we have to consider all possible subsets in $\mathcal X$. However, it will be computationally cumbersome to enforce the full influence-limiting policy on a cooperative spectrum sensing network with a large number of nodes, as the total number of subsets in  $\mathcal X$ with $n$ nodes is $\sum^{n}_{k=1} {n \choose k}$.

We provide a practical strategy to reduce the complexity of the influence-limiting policy. We introduce a parameter $\eta, 0 < \eta \leq n$, in the influence-limiting policy and consider limiting the influence of any subset which has no more than $\eta$ nodes. For example, $\eta=1$ means that we perform the influence-limiting policy only on each individual node inside the network and $\eta = n$ means a full scale influence-limiting policy evaluated on all possible combinations of $\mathcal X_{\text{sub}}$. By adjusting the value of $\eta$, we can balance the complexity of the influence-limiting policy and the effectiveness against potential attacks. 

\begin{algorithm}[t]
\SetAlgoLined
\SetKwInOut{Input}{Input}\SetKwInOut{Output}{Output}
\Input{Historical sensing results $\mathcal{X}$ and the corresponding decision outputs in $\mathcal Y$; parameters $c_1,c_2,\eta$.}
 \BlankLine
 \textbf{for each $\mathcal X_{\text{sub}}$ according to the value of $\eta$}:\\
 ~~~~ ~~~~Calculate the decision-flipping influence \\
 ~~~~ ~~~~$I(\mathcal X_{\text{sub}})$;\\
 ~~~~ ~~~~Compute the threshold $\delta(|\mathcal X_{\text{sub}}|)$;\\
 ~~~~ ~~~~Enforce influence through limiting the weight \\
 ~~~~ ~~~~$w(\mathcal{X}_{\text{sub}})$;\\
 \textbf{end for}\\
Feedback $w(\mathcal{X}_{\text{sub}})$ to traditional/existing defenses.\\
\caption{Influence-limiting defense}\label{Alg:Search}
\label{algorithm:defense}
\end{algorithm}

The step-by-step process of the influence-limiting defense can be seen in Algorithm \ref{algorithm:defense}. The advantage of using influence-limiting policy is that we can add this enforcement as a parallel, non-invasive constraint to the primary/traditional decision method at the fusion center. The influence-limiting policy is applied to bridge the gap between traditional defenses and the new adversarial machine learning-based attacks, such as LEB attacks.

\subsection{Experimental Validation}
We evaluated the performance of the proposed influence-limiting defense using the collected dataset. Experimental configuration is the same as that in Section~\ref{exper_eval}. In this part of experiments, we adopted the trust value-based method as the existing fusion center defense mechanism due to its popularity in cooperative spectrum sensing \cite{rawat2011collaborative}.

\subsubsection{Impact of $c_1$ and $c_2$ for $\delta(|\mathcal X_{\text{sub}}|)$}

We first measured the impact of parameters $c_1$ and  $c_2$ with regard to the threshold $\delta(|\mathcal X_{\text{sub}}|)$ in Eq. \eqref{eq:threshold} on the defense performance of the influence-limiting defense.

It is obvious that the limitation caused by  $\delta(|\mathcal X_{\text{sub}}|)$ will lead to a performance cost for the fusion center, especially when there is no malicious node. The cost essentially depends on $c_1$ in Eq. \eqref{eq:threshold}. According to the mathematical property of Sigmoid function, less limitation will be enforced when $c_1$ is smaller; i.e., the cost when no malicious nodes are present will be smaller.

On the other hand, a smaller $c_1$ hurts the system more when malicious nodes are indeed present. Similarly, a larger value of $c_2$ makes sure that a potentially malicious node will be penalized more but may also penalize legitimate nodes more when there is no attack. Therefore, we evaluated a wide range of choices of $c_1$ and $c_2$ to observe the balance between the performance without the attack and the defense effectiveness in the presence of attacks.

\begin{figure}[t]
\centering
\includegraphics[scale=01]{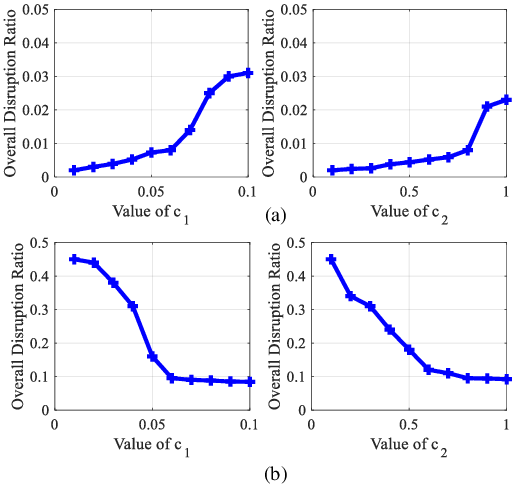}
\caption{The relationship between different values of parameters $c_1$ and $c_2$, and the overall disruption ratios (a) when no malicious node exists and (b) when LEB attacks are present.}
\label{Fig:before_parameters}
\end{figure}

\textit{Case 1: No malicious node.}
When no malicious node exists, all the 20 sensing nodes report the true sensed values to the fusion center. We illustrate the overall disruption ratio in 1000 timeslots with regard to influence-limiting defense of different values for $c_1$ and $c_2$ in Fig.~\ref{Fig:before_parameters}(a). When evaluating one parameter, we fixed the other parameter as the two parameters are independent from each other. We observe from Fig.~\ref{Fig:before_parameters}(a) that when no malicious node is present, the overall disruption ratio increases as both $c_1$ and $c_2$ increase, which corresponds to the slight cost of the influence-limiting defense.



\textit{Case 2: With LEB attacks.}
When LEB attacks are present, we evaluated the scenario where 8 out of 20 nodes are malicious (i.e., $m=8$ and $n=20$). The relationship between different values of $c_1$ and $c_2$ and the overall disruption ratio is shown in Fig.~\ref{Fig:before_parameters}(b). When $c_1$ or $c_2$ is very small, the influence limitation is negligible. The defense performance is demonstrated when $c_1$ and $c_2$ increase beyond 0.05 and 0.5, respectively.

To make the influence-limiting defense achieve the full potential of defense capability, we choose  $c_1$ and $c_2$ to decrease the overall disruption ratio when malicious nodes exist, while still maintaining a slight cost when no malicious node is present. Combining the experimental results of the above two cases, we can observe that $c_1$ and $c_2$ can be set as 0.05 and 0.5, respectively, in our example.

\subsubsection{Impact of $\eta$ on defense}
\begin{figure}[t]
\centering
\includegraphics[scale=0.5]{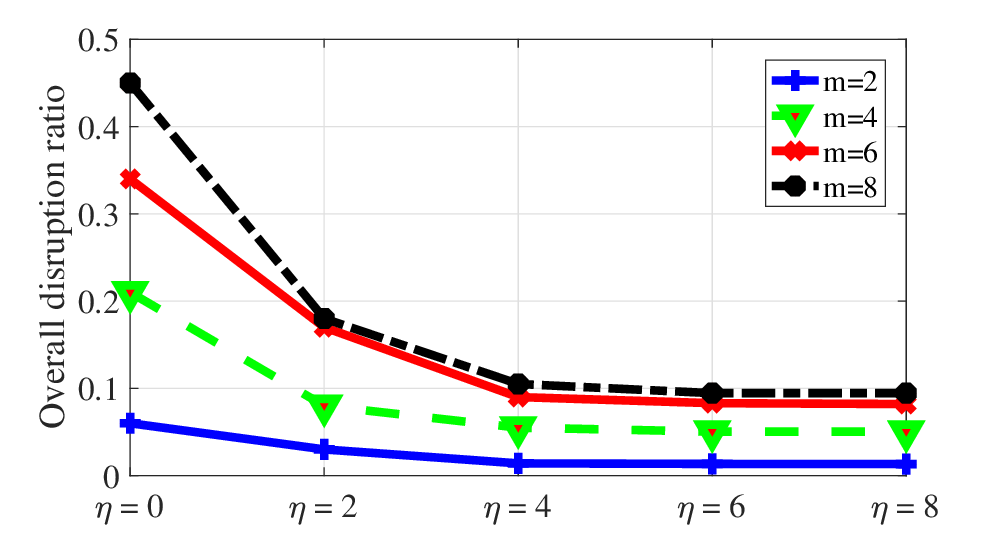}
\caption{The overall disruption ratios given different value of $\eta$ with different $m$.}
\label{Fig:eta}
\end{figure}

We also evaluated the defense performance under different values of $\eta$ that balances the complexity and the defense performance of the influence-limiting policy. We assume $c_1 = 0.5, c_2 = 0.05$ for these experiments. The overall disruption ratio with LEB attacks is shown in Fig.~\ref{Fig:eta} when we vary the number of malicious nodes. The results demonstrate that when $\eta$ is equal or larger than 4, the performance improvement is almost negligible for different cases of $m=2, 4, 6$ and $ 8$.

\subsubsection{Varying the number of malicious nodes}
Given the parameters $c_1$, $c_2$ and $\eta$, we compared the influence-limiting defense with existing defenses in cooperative spectrum sensing.

We first compared the defense performance of four cases in cooperative spectrum sensing: (i) no LEB attack and no influence-limiting defense at the fusion center; (ii) LEB attacks are present without the influence-limiting defense; (iii) No LEB attack while influence-limiting defense is present; (iv) both LEB attacks and influence-limiting defense are present.

We compared the performance in terms of the overall disruption ratio for $m=2, 4, 6, 8$ in Table~\ref{table:four_cases}, which demonstrates that the cost of influence-limiting defense is slight when no malicious node is present (with an overall disruption ratio of 0.008 compared to 0.002 when influence-limiting defense is absent).  Compared with the case when no influence-limiting defense is applied, the average defense performance improvement of influence-limiting defense in terms of overall disruption ratio is around 78\% when LEB attacks exist, which shows the effectiveness of the defense.

\begin{table}[t]
\caption{The overall disruption ratios with different numbers of malicious nodes $m$.}
\centering
        \begin{tabular}{p{1.5cm}p{2cm}|p{0.6cm}p{0.6cm}p{0.6cm}p{0.6cm}}
        \hline
       \multirow{2}{*}{\textbf{LEB attacks}} & \multirow{1}{*}{\quad \textbf{Influence-}} &\multicolumn{4}{c}{\textbf{$m$}}   \\ \cline{3-6} &  \multirow{1}{*}{\textbf{limiting defense}}  & \quad \textbf{2} & \quad \textbf{4} & \quad \textbf{6}  & \quad \textbf{8} \\ \hline
        \quad  absent & \quad \quad  absent          &  ~ & ~~~~~0.002    & ~     &  ~  \\ \hline
        \quad present & \quad \quad    absent         & 0.06  & 0.21  & 0.34   & 0.45  \\ \hline
        \quad  absent  &  \quad \quad   present        &  ~ & ~~~~~0.008    & ~     &  ~  \\ \hline
        \quad  present & \quad \quad    present        & 0.013 & 0.05     & 0.082     & 0.095   \\
        \hline \\
    \end{tabular}
\label{table:four_cases}
\end{table}

\subsubsection{Performance comparison with existing defenses}
\begin{figure}[t]
\centering
\includegraphics[scale=0.45]{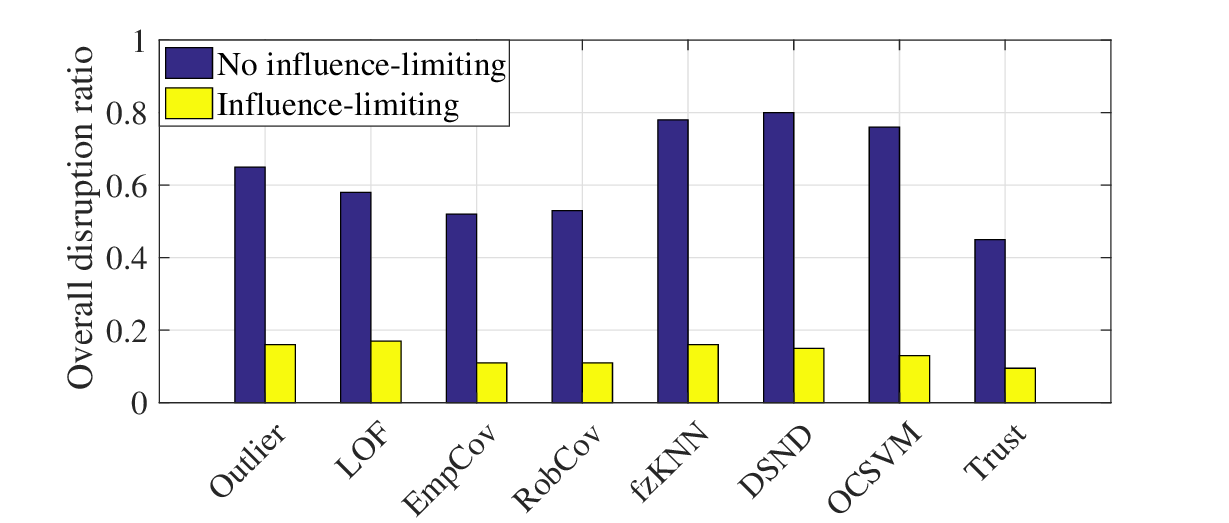}
\caption{The overall disruption ratio comparison of influence-limiting policy for different defenses.}
\label{Fig:in_def_comp}
\end{figure}

Next, we compare the performance of influence-limiting defense with other existing defenses in terms of overall disruption ratio. We employed the same configurations for parameters of influence-limiting defense and other defenses, and set $m = 8$. The performance illustrated in Fig. \ref{Fig:in_def_comp} validates our proposed defense by dramatically decreasing the overall disruption ratio by around 80\% on average.

The above defenses are general intrusion detection methods used to detect outliers. We also compared influence-limiting defense with specialized defense methods designed specifically for cooperative spectrum sensing. Double-Sided Neighbor Distance (DSND) \cite{li2010catch}  is a malicious nodes detection method based on the assumption that \textit{if a sensing node's history is too far away or too close to other's histories, then it might be abnormal and is probably a malicious node.} A malicious node identification method and an adaptive linear combination rule were provided in  \cite{chen2017cooperative}. Besides defenses of IMBA, a modified Combinatorial Optimization Identification (COI) algorithm was proposed in \cite{qin2013defending} to deal with cooperative attacks. An adaptive reputation-based clustering algorithm was presented in \cite{hyder2011defense} to defend against both IBMA and CMBA.

We quantified defense performance in terms of overall disruption ratio to compare influence-limiting defense with existing defenses in cooperative spectrum sensing. LEB attacks are deployed as the attack method and the malicious nodes are set as $m=8$. The fusion center is configured with SVM as the decision rule. Unlike previous experiments where we run the influence-limiting defense with other defenses, here we run the influence-limiting defense as the only defense mechanism at the fusion center. The detailed performance is shown in Table \ref{Tab:diff_defenses}.

The performance of influence-limiting defense is slightly better than existing methods even run independently. LEB attacks have better attack performance on reputation-based defenses like method in \cite{hyder2011defense}, which ends up with 38\% overall disruption ratio, than other methods.
\begin{table}[t]
\begin{center}
\centering
\caption{Overall disruption ratio (\%) comparison under different defense methods.}\label{Tab:diff_defenses}
\begin{tabular}{c|c|c}
\hline
\textbf{Defense methods} & \textbf{Without defense} & \textbf{With defense}  \\
\hline
Li et al.\cite{li2010catch}    & 85 & 36  \\
\hline
Chen et al. \cite{chen2017cooperative}    & 85 & 33   \\
\hline
Qin et al. \cite{qin2013defending}   & 85 &  31 \\
\hline
Hyder et al. \cite{hyder2011defense} & 85 &  38 \\
\hline
Influence-limiting defense    & 85 &  27 \\
\hline
\end{tabular}
\end{center}
\end{table}

\textbf{Summary: }The experiment results above demonstrate the robust performance of the proposed influence-limiting defense. Influence-limiting defense can defend against both independent Byzantine attacks and cooperative Byzantine attacks. Similar to many security mechanisms, there will be a performance cost especially when no malicious nodes are present. How to better balance the security and performance is identified as potential future work.

\section{Limitations and Discussion} \label{sec:limitations}
Although our proposed LEB attack framework and influence-limiting defense achieved promising attack and defense performance, it is important to note the cost of the fusion center. We conducted experiments from different aspects to validate the LEB attack framework. However, it can not guarantee attack performance. The LEB attack framework is based on machine learning techniques, which still have many unanswered questions \cite{papernot2016towards,papernot2016limitations,shokri2017membership}.  We list limitations of the LEB attack framework as follows:
\begin{itemize}
\item How to choose the minimum number of and optimal sub-models in the surrogate model depends on specific scenarios. In the design, we make the sub-model as a flexible algorithm set such that new models can be added and current sub-models can be removed. But how to decide the number and types of sub-models in the surrogate model in practice is left as a design choice, which we identify as potential future work.
\item We need to train and update all the sub-models concurrently in the training and testing process. In our experiments, the average convergence time for most sub-models is approximately 500 timeslots. Although we have employed incremental/online learning techniques to minimize the update cost, a better solution still lies ahead as potential future exploration.
\item Experimental results of the LEB attack framework are data-dependent. i.e., the attack performance is empirical and might change for different datasets. How to prove the effectiveness of the LEB attack framework from a theoretical perspective is also identified as potential future work.
\end{itemize}

\hlb{From a broader AI perspective, our proposed LEB attack framework  can be applied to many other scenarios that require data fusion process such as in Internet of Things (IoT) \cite{ding2019survey}. Another important application is in machine learning itself. Machine learning-based attack framework does not necessarily  provide explainability. For example, machine learning models like deep neural networks are treated more like a black box without giving much transparency and accountability to the users. Machine learning models are usually data-driven \cite{papernot2016towards,papernot2016limitations}, thus the performance is usually empirical and highly dependent on the given datasets. In our proposed LEB attack framework, we employ an algorithm set to counter the uncertainty such that the attack utility has higher probability to achieve its goal. Besides, we employ the algorithm set to mutually validate the malicious inputs (malicious samples will be taken as inputs for each sub-model).}

\hlb{From the defense perspective, the influence-limiting defense is effective in limiting the damage caused by attacks. It can be also applied to a broader machine learning scenario. However, as shown in experimental results, it will also lead to performance cost when there is no attack. Besides, to enforce a full version of the influence-limiting defense, the computation cost still needs to be addressed. In this paper, we explored the defense from a direct way, which is to limit the influence of sensing nodes. Our design of $\delta(|\mathcal{X}_{sub}|)$ provides a feasible and effective trade off between performance and complexity, and we found that in practice it is not necessary to enforce a full-version of the influence-limiting defence to achieve a satisfactory performance.}

\section{Related Work} \label{sec:related}
\textbf{Cooperative spectrum sensing and defense} Cooperative spectrum sensing is an efficient way to detect spectrum usages of TV white space channels. Defenses like statistics-based, machine learning-based, and trust-based methods have been widely studied. Statistics-based defenses assign different statistics to sensing nodes. For example, an outlier is computed for each node in \cite{kaligineedi2010malicious}; \cite{chen2017cooperative} obtains weighted coefficients of each node; and \cite{chen2012robust} proposes a method through majority vote of neighboring nodes, a Bayesian method is discussed in \cite{penna2012detecting}. Machine learning-based methods leverage machine learning techniques to classify legitimate and malicious data. For example, \cite{fatemieh2011using,rajasegarar2015pattern} utilize supervised learning based on SVM to distinguish malicious nodes; \cite{li2010catch} uses different KNN-based algorithms in an unsupervised way to detect malicious nodes; and \cite{thilina2013machine,wang2018primary} discuss both supervised and unsupervised ways to defend the fusion center. A trust-based method \cite{chen2008robust,rawat2011collaborative, sagduyu2014trust} maintains a trust value for each node, which will be used as a weight in the global decision process.

Taking into account such a wide range of defenses, we present a powerful attack mechanism based on the LEB attack framework, which aims to learn the defense at the fusion center and then launch effective attacks. Based on the proposed attack strategy, we introduce an influence-limiting defense, which is a non-invasive method that can coexist with existing defenses to bridge the gap between traditional defenses and new LEB attacks.

\hlb{There are some existing studies that deal with smart/intelligent attackers in cooperative spectrum sensing. In \cite{duan2012attack}, a smart attack framework that can maximize the expected aggregated reward of the attacker was proposed. In \cite{duan2012attack,wang2018trust}, smart defense mechanisms were proposed to discourage attackers. Our work differs from these studies in that the LEB attack framework focuses on using a black-box strategy to minimize the attacker's trace (via LEB) and the malicious perturbation when the attacker decides to launch the attack.}

\textbf{Adversarial machine learning} Adversarial machine learning focuses on learning under the existence of active adversaries \cite{hayes2017machine}. The transferability \cite{papernot2016transferability} in machine learning models gives adversaries the opportunity to learn and compromise targeted models. Adversarial example generation methods, such as iterative methods \cite{kurakin2016adversarial, moosavi2016deepfool} and gradient-based methods \cite{goodfellow2014explaining}, are well-known ways to create malicious data targeting a machine learning classifier under specific scenarios. Our proposed adversarial sensing data generation algorithm achieves similar performance as other existing methods, while reducing by 65\% the computational cost on average.

\hlb{There is a growing interest in applying adversarial machine learning to wireless communications \cite{Sagduyu2020}. 
    Exploratory (inference) attacks were studied in \cite{shi2018inference, erpek2019inference} to build a surrogate model (to mimic wireless transmission patterns) that is used to develop a smart jamming scheme.
Causative (poisoning) attacks were developed to manipulate the training data input to machine learning classifiers \cite{shi2018causative, sagduyu2019causative, Luo2020IoT} with applications in wireless access and IoT. A similar approach was followed in \cite{kemal2019trojan} to design a Trojan attack, where the adversary adds Trojans (triggers) to the training data and activate them later in test time against a wireless signal classifier. On the other hand, membership inference attacks were considered in  \cite{Shi2020MIA} to learn whether a particular data sample has been used to train a wireless signal classifier. Finally, evasion attacks were studied in \cite{sagduyu2019evasion}, where an adversary jams the sensing period to change the inputs to a machine learning classifier and force a target transmitter into making wrong transmit decisions. Evasion attacks were also considered to fool modulation classifiers \cite{Sadeghi2019, Bair2019, Kokalj2019, Gunduz2019, Kim1, Kim2, Kim3}. Compared to previous works, this paper considers adversarial machine learning attacks against cooperative spectrum sensing in wireless communications and presents corresponding defenses.}

\section{Conclusion} \label{sec:conclusion}
In this paper, we explored the partial model problem (i.e., the attacker controls part of the input dimensions) in cooperative spectrum sensing. Our results raise the need to revisit the attack and defense spaces for cooperative spectrum sensing as new learning-based attack models have emerged. Inspired from the model mismatch phenomenon in real world, we proposed a powerful attack mechanism named the LEB attack against cooperative spectrum sensing, which offers an effective attack paradigm against cooperative spectrum sensing. The LEB attack is designed in a flexible and generic way and can be applied to different attack scenarios. We used comprehensive experiments to show the severe impact of LEB attacks on the cooperative spectrum sensing. Given the gap between existing defenses and new adversarial learning-based attacks, such as LEB attacks, we designed a make-up mechanism, named influence-limiting defense, sided with existing defenses to counter LEB attacks and other similar attacks. We showed that influence-limiting defense is successful in reducing the overall disruption ratio of malicious nodes.

\bibliographystyle{IEEEtran}
\bibliography{reference}
\end{document}